\documentclass[manuscript,   onecolumn]{aa}
\usepackage{graphicx}
\usepackage{aas_macros}
\usepackage{natbib}
\usepackage{amsmath}
\usepackage{amssymb}

\bibpunct{(}{)}{;}{a}{}{,}

\begin{document}

\title{An analytical approach to the dwarf galaxies cusp problem}

\author{R. S. de Souza\inst{1,2}   \and E. E. O. Ishida\inst{1,2}}
\offprints{Rafael S. de Souza \email{rafael.desouza@ipmu.jp}}

\institute{$^1$IAG, Universidade de S\~{a}o Paulo, Rua do Mat\~{a}o 1226, Cidade
Universit\'{a}ria, \\ CEP 05508-900, S\~{a}o Paulo, SP, Brazil\\
$^2$IPMU,
The University of Tokyo,
5-1-5 Kashiwanoha, Kashiwa, 277-8583, Japan}

 \date{Accepted -- Received}

\date{Released Xxxxx XX}

\authorrunning{de Souza \& Ishida}
   \titlerunning{The dwarf galaxies cusp problem}

\abstract
{}
{An analytical solution for the discrepancy between observed core-like profiles and predicted cusp  profiles in dark matter halos is studied.}
{ We calculate the distribution function for Navarro-Frenk-White halos and extract energy from the  distribution,  taking into account the effects of baryonic physics  processes.}
{We show with  a simple argument  that we can reproduce the evolution of a cusp to a flat density profile by a  decrease of the initial potential energy.}
{}
\maketitle

\section{Introduction}

There are many dwarf galaxies in the Universe (e.g.~\citet{mar97,saw10}), and the Local Group is an excellent place to study these systems and their properties \citep{mat98,pas10}.

 The  nearly solid body rotation curve  observed in dwarf galaxies indicates  a central core in the dark matter distribution  (e.g., \citet{bur95, deb02, gen05,deb05}). These results disagree with predictions from numerical simulations, which require density profiles with cusps  (e.g. \citet{nav97,mor99,nav04}).

\citet{nav96} were one of the first to point out the possibility that feedback mechanisms can turn the central dark-matter cusp into a cored one.
Following this line of investigation, \citet{gneezao} analyzed the influence of winds.
\citet{rea05} discussed  how  external impulsive mass-loss events can successfully act as a  flattener for the central density cusps,  and
\citet{mas06} showed that random bulk motions of gas in small primordial galaxies could flatten the central dark matter cusp in $\sim 10^{8}$ years.  Beyond that, several authors have suggested that the interstellar medium (ISM) of dwarf galaxies  systems could be entirely removed as a result of star formation \citep{dek86,mor97,mur99,mac99,mor02,hen04, mor04}.

In the present work, we construct a simple model under the hypothesis that astrophysical processes are able to remove baryonic  gas from dwarf galaxies. We show for the first time a clear  and analytical argument to solve the cusp problem.  We argue  that baryonic feedback  could change the distribution function on the phase space of energy and angular momentum,  flattening the density profile.

This paper is organized as follows. In Sect.  2, we briefly discuss the removal of baryonic gas from dark matter halos as a consequence of different astrophysical processes. In Sect.  3,  we review the formalism of distribution functions. In Sect. 4,  we discuss the changes in density profile caused by gas removal and show our results. Finally   we present  our conclusions in Sect. 5.

\section{Expulsion of baryonic gas}	

Our main goal is to give an analytical explanation to the flat density profile observed in dwarf galaxies. Our starting point will be the results obtained from $\Lambda$CDM numerical simulations \citep{mor99,nav04}. In this context, we begin with a dwarf galaxy that has a Navarro-Frenk-White profile (NFW) \citep{nav97}  after the formation of its first stellar population, and study the signatures left in the dark matter profile when part of its baryons are removed. The first natural question is: why are the baryons removed?

Before answering that question, it is important to emphasize that  a cusp profile is not only characteristic, but necessary.  In the initial phase of galaxy formation, when only a small part of the stellar population is formed, a deep gravitational potential well is needed in order to trap the baryons \citep{mor97,bra98}. Beyond that, baryon accretion in the center of the halo will suffer adiabatic contraction and drag the dark matter profile into an even steeper one \citep{blu86}. 

On the other hand, the baryonic characteristic of the gas causes it to radiate energy, increasing the entropy of the dark matter background and leaving strong signatures in the density profile \citep{elz01,rom08}. In this context,  also the  effects of gas bulk motions need to be considered, as shown in \cite{mo02} and \cite{mas06}.

After this initial collapse, the galaxy will go through one (or probably more) star-formation bursts. \citet{mac99} and \citet{pas10}  showed that SF alone can remove almost all the interstellar baryonic gas inside dwarf galaxies, for $M_{halo}\leq 10^7 M_{\odot}$.  \citet{gelato99} explored how different mass ejection fractions and disk density distribution could fit the observed rotation curve of the dwarf galaxy  DDO 154.
 \citet{hay03} demonstrated that the mass loss can vary from $5\%$ to $95\%$ as a result of tidal effects  caused by  a host larger galaxy.

Below we will consider only astrophysical effects that  are able to remove from a small percentage up to all the baryonic gas initially trapped by the dark matter halo. Following \citet{pas10}, we estimate that $1/50$ of the total mass in dwarf galaxies is visible. Moreover, $15\%$ of this visible mass is in form of stars and $85\%$ is in form of interstellar gas. We consider that the astrophysical processes cited before are able to remove from $5\%$ to $100\%$ of the gas in the galaxy halo. This means that for halos with $M_{halo}\sim 10^7 M_{\odot}$, from $0.085\%$ to $1.7\%$ of the total halo mass can be removed.

Note that the crucial point in our analysis is the percentage of total halo mass lost. All the other astrophysical mechanism cited above that do not necessarily remove baryons are not considered here. Our results are valid for any mass loss mechanisms, as long as they expel the same percentage of baryons out of the dark matter halo potential well.

\section{Dark matter density profile}

At a given time, $t$, a full description of any collisionless system  can be obtained by specifying the number of particles, $f(x,v,t)d^{3}xd^{3}v$,  that have  positions in the small volume $d^{3}x$ centered in $x$ and small range of velocities  $d^{3}v$ centered in $v$. $f(x,v,t)$ is called the distribution function, hereafter DF \citep{bin08}. The DF of a  mass distribution in a steady state is related to its density profile through
\begin{equation}
\rho(r) = \int f(r,v)d^{3}v\text{ .}
\label{rho1}
\end{equation}

For any general spherical system, we can define a  DF,  $f(E,L)$,  which depends on the energy, $E$, and the angular momentum absolute value, $L$.  \citet{osi79} and \citet{mer85} found that in anisotropic spherical systems, $f$ can be expressed as:
\begin{equation}
f(Q,L^{2}) = \mathcal{F}\left(Q\right)L^{2\alpha},
\label{fQL2}
\end{equation}
where $\alpha$ is  a real number and

\begin{equation}
Q = E-\frac{L^{2}}{2r^{2}_{a}}\text{,}
\end{equation}
for a radial parameter $r_{a}$, known as the anisotropy radius. It is useful to define the relative potential, $\Psi=-\phi$, where $\phi$ is the gravitational potential.

For distribution functions that can be written in the form of Eq. (\ref{fQL2}), the fundamental integral equation can be written  as \citep{cud91,bin08}

\begin{align}
\rho(r) &= \frac{2\pi}{r^{2}}\int^{\Psi}_{0}\mathcal{F}(Q)\,dQ \,\, \\
\times& \int^{2r^{2}(\Psi-Q)/(1+\frac{r^{2}}{r^{2}_{a}})}_{0}\frac{L^{2\alpha}\,dL^{2}}{\sqrt{2(\Psi-Q)-(L^{2}/r^{2})(1+\frac{r^{2}}{r^{2}_{\alpha}})}}.\nonumber&
\end{align}
After performing the integration over $L$, we obtain a single integral equation for $\mathcal{F}(Q)$,
\begin{equation}
\rho(\Psi)=\Theta(\alpha) \int_0^{\Psi}\left( \Psi-Q \right)^{\alpha+1/2}\mathcal{F}(Q)\mathrm{d}Q~,\label{rhopsi1}
\end{equation}
where
\begin{equation}
 \Theta(\alpha) = 2^{\alpha+3/2} \pi^{3/2}\frac{\Gamma(\alpha+1)}{\Gamma(\alpha+3/2)}r^{2\alpha}{\left(1+\frac{r^2}{r^2_\alpha}\right)}^{1-\alpha}.
\end{equation}
For a general $\alpha$, Eq.  (\ref{rhopsi1}) can be inverted, resulting in
\begin{equation}
\mathcal{F}(Q) = \frac{\sin(n-1-\alpha)\pi}{\pi\Theta(\alpha)(\alpha+1/2)!}\frac{d^{n+1}}{dQ^{n+1}}\int^Q_0\frac{\rho(\Psi)}{(Q-\Psi)^{\alpha+\frac{3}{2}-n}}\mathrm{d}\Psi,
\label{foq}
\end{equation}
with $n$ defined by
\begin{equation}
n\equiv \left[\alpha+1/2\right]+1,
\end{equation}
and $\left[\alpha+1/2\right]$ representing the largest integer less than or equal to $(\alpha+1/2)$.

From the above expressions, it is clear that any perturbation in $E$ and/or $L$ will change the DF (Eq.  (\ref{fQL2})) and, consequently, leave a signature in the galaxy density profile (Eq.  (\ref{rho1})). Our strategy in this work is to start with a NFW profile in Eq.  (\ref{foq}), find it a suitable DF and analyze the change in the density profile caused by perturbations in the DF. Although, as we initially proposed an entirely analytical calculation, we must express the NFW profile in a way that makes  a completely analytical approach possible. This is done in the next section.

\section{Erasing the  cusp}
\subsection{Spherical density profiles}
In order to evaluate the distribution function for a cusp-like profile, we need to consider an initial density profile with a central cusp and place it into Eq.  (\ref{foq}). For simplicity, we adopt the following family of spherical potentials \citep{rin09},
\begin{equation}
\phi(r) = - \frac{\phi_{0}}{\left(1+\left(\frac{r}{b}\right)^{\eta}\right)^{\gamma}},
\end{equation}
where $\phi_{0}$ is the value of  potential inside the halo radius, $\eta, \gamma$,  and $b$ are positive parameters.
Using the Poisson equation, we found the dimensionless density profile corresponding to  this potential,
\begin{equation}
\widetilde{\rho}(r) = \frac{b^{\eta\gamma+2}}{1+\eta} \frac{(1+\eta)b^{\eta}+(1-\eta\gamma)r^{\eta}}{r^{2-\eta}(b^{\eta}+r^{\eta})^{\gamma+2}}.
\label{rhoad}
\end{equation}

Defining $\widetilde{\Psi}\equiv-\phi(r)/\phi_0$,
we can also write the density as function of the relative potential,
\begin{eqnarray}
\widetilde{\rho}(\widetilde{\Psi})&=&\frac{1}{\eta +1}\left\{\left[(1-\eta  \gamma ) \widetilde{\Psi} ^{\frac{1}{\gamma }+1}+\eta  (\gamma
   +1) \widetilde{\Psi} ^{\frac{2}{\gamma }+1}\right] \times \nonumber\right.\\
   & &\times \left.\left(\widetilde{\Psi} ^{-1/\gamma
   }-1\right)^{1-\frac{2}{\eta }}\right\},\label{rhopsi}
\end{eqnarray}
where the tilde denotes dimensionless quantities (for more details see  \citet{rin09}).  The Eq.  (\ref{rhopsi}) is general enough to be applied to any spherical density profile, given an appropriate choice of parameters. The advantage of this expression is that it permit us to  write $\widetilde{\rho}$ in terms of $\widetilde{\Psi}$, which is not always analytically possible. However,  we can always  find a fit of Eq.  (\ref{rhopsi}) corresponding to a specific density profile.

\subsection{Connection to the Navarro-Frenk-White profile}
One of the strongest predictions from cosmological simulations is the universal density profile present in dark matter halos \citep{nav97},
\begin{equation}
\rho(r) = \frac{\rho_s}{r/rs(1+r/rs)^{2}},
\end{equation}
where $\rho_s$ is de central density  and $r_s$ is the scale radius of the halo.

\begin{figure}
  \includegraphics[width=\columnwidth]{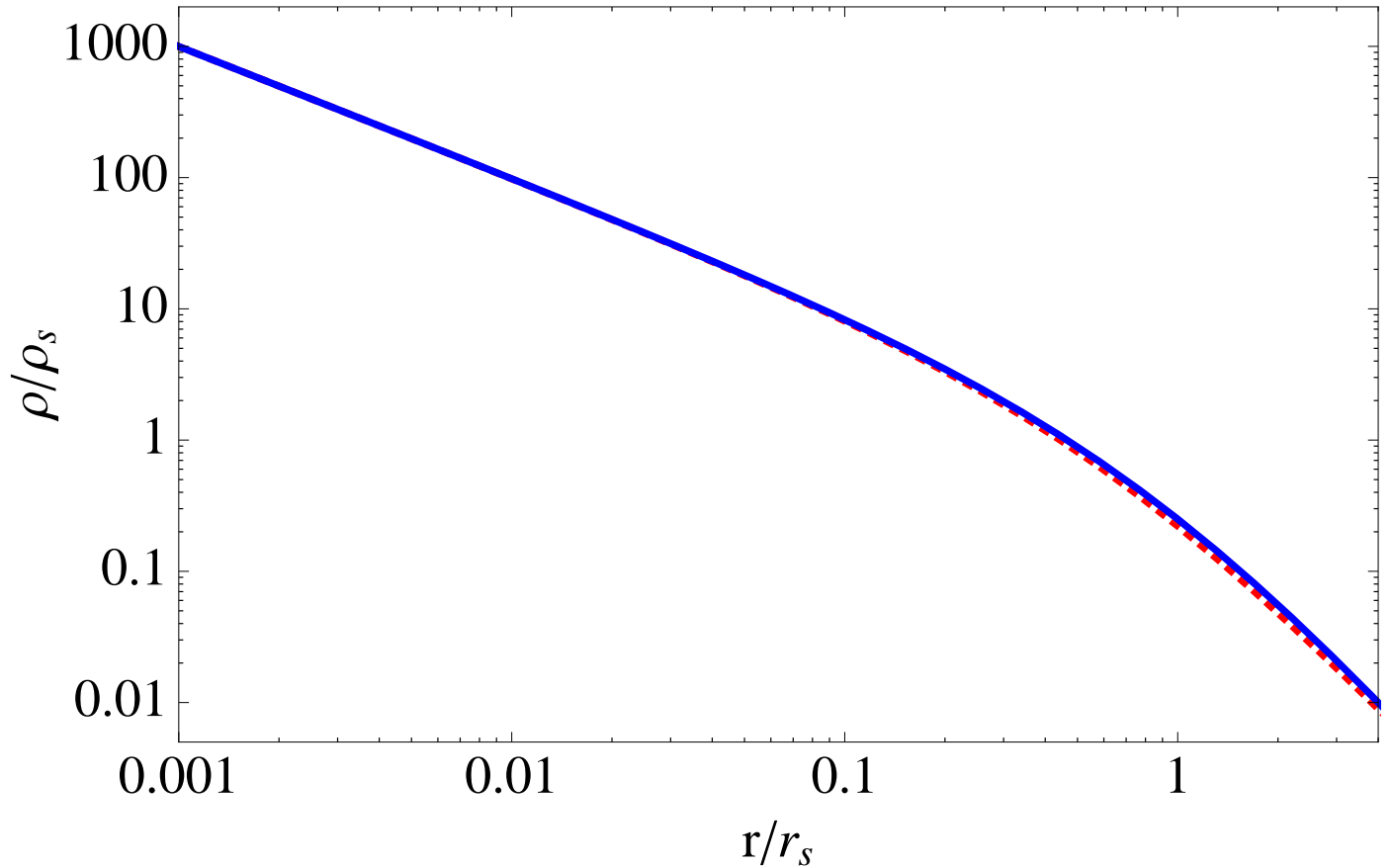}
 \caption{Reconstructed density profile (red dashed line) compared with original NFW profile (blue solid line).}
 \label{fig:one}
\end{figure}

Using  $\eta =1 , b = 1$ and $\gamma = 1/2$, Eq.  (\ref{rhoad}) becomes numerically similar to NFW profile, as we can see in Fig. \ref{fig:one}.
Given this choice of parameters, $\widetilde{\Psi}$, Eqs. (\ref{rhoad}) and (\ref{rhopsi}) become

\begin{equation}
\widetilde{\Psi}(r) = \frac{1}{\sqrt{r+1}},
\end{equation}

\begin{equation}
\widetilde{\rho}(r) = \frac{r+4}{4 r (r+1)^{5/2}},
\label{rhoajus}
\end{equation}

and
\begin{equation}
\widetilde{\rho}(\widetilde{\Psi}) = \frac{3 \widetilde{\Psi} ^7+\widetilde{\Psi} ^5}{4-4 \widetilde{\Psi} ^2}.
\label{tiorho}
\end{equation}

\begin{figure}
  \includegraphics[width=\columnwidth]{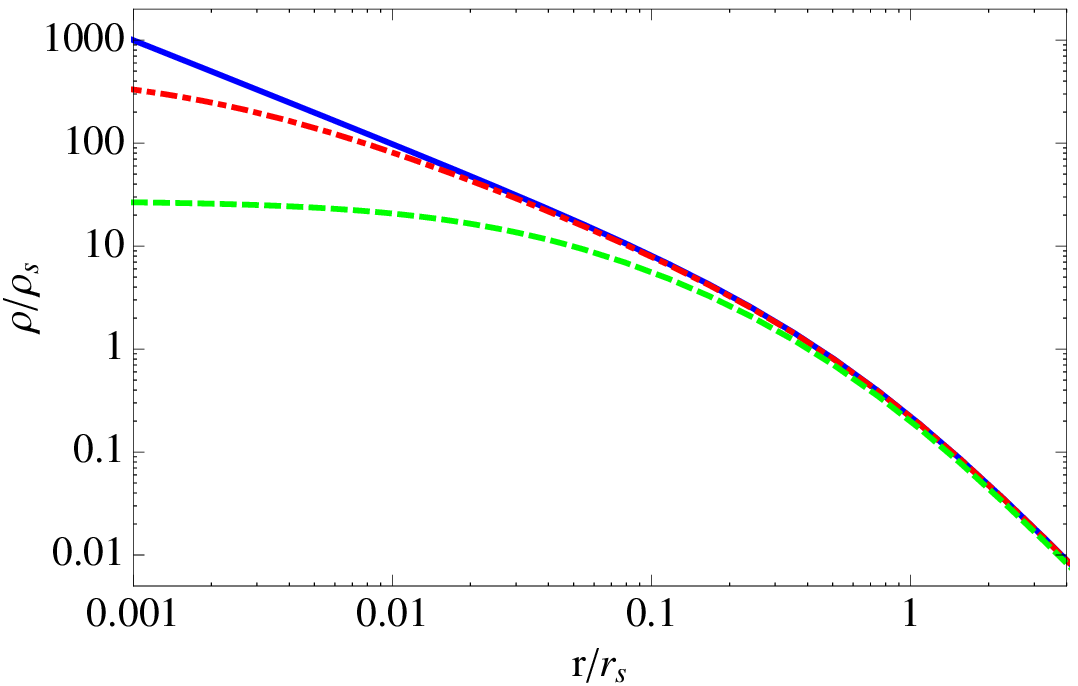}
 \caption{New equilibrium configuration of the perturbed halo removing 0.1\% (red dot-dashed line) and 1.7\% (dashed green line) of total energy Q,  compared to the NFW profile (blue solid line).}
 \label{fig:two}
\end{figure}

In order to test our calculations, we first found a DF for the NFW profile and reconstructed the density profile using  the Eq. (\ref{tiorho}) in  Eq.  (\ref{foq}). We found that  for $\alpha \sim 1/2$ and $r_{\alpha} \sim 1$,  Eq.  (\ref{foq}) agrees well with the DF for NFW profile.  With these parameter values we have

\begin{equation}
\mathcal{F}(Q)=-\frac{Q^3 \left(15 Q^6-39 Q^4+27 Q^2+5\right)}{2 \pi ^2 \left(Q^2-1\right)^3}\Theta(1/2)^{-1}.
\end{equation}
The above expression placed into Eq.  (\ref{rhopsi1}) recovers Eq.  (\ref{tiorho}).

We  suppose that the macroscopic variables, like radial density profile, will change if the halo experiments a physical process that perturbs its DF. We considered that  the removal of baryonic gas will shift the energy of the system by $Q\rightarrow Q\pm\Delta Q$. According to the arguments in Sect.  2, SF and external tidal effects can  remove $(0.085-1.7)\%$ of the halo gravitational potential energy.  In order to find a new DF configuration,  we have to evaluate $\mathcal{F}(Q-\Delta Q) \approx \mathcal{F}(Q)- \Delta Q\frac{\partial \mathcal{F}}{\partial Q}$.  Which is just a solution of the Boltzmann equation, $\frac{\partial \mathcal{F}}{\partial t}+\frac{\partial Q }{\partial t}\frac{\partial \mathcal{F}}{\partial Q}=0$, if we consider that the gas removal occurs in a short time interval, $\Delta t \approx dt$.

The results are presented in Fig. \ref{fig:two}, where we show the new configuration of the  density profile after remotion of $(0.085-1.7)\%$ from gravitational potential energy. It is evident from this figure that a little change in the total energy of halo can flatter the profile in the inner region. Note that the core-like behavior we were looking for is already evident for $\Delta Q \approx 0.001$. As a consequence, any feedback capable of removing $\approx 10\%$ of interstellar  gas would be able to smooth the matter density profile.

\section{Conclusions}

The N-body simulations based on collisionless cold dark matter do not show the core-like behavior observed in local dwarf galaxies. Instead, they are better described by a steep power-law mass density distribution called cusp. Several attempts were made with numerical simulations  in order to take into account the gas dynamics in the evolution of dark matter profiles.

Considering that the total energy of the halos will change because of  baryonic feedback, we showed that it is possible to model the evolution of an initial cusp-like profile into a core-like one. We used an analytical analysis based on the density profile DF.

The advantage of this approach is that we obtain results similar to those extracted from numerical simulations,  which take into account the gas dynamics \citep{pas10,gov09} in a very simple way. On top of that, considering the baryon loss as a mechanism of flattening the dark matter density profile of dwarf galaxies allows us to support the low amount of baryons observed in dwarf spheroidal galaxies \citep{gil2007}.

However, as discussed in the introduction, there are two  competitive processes occurring during the formation of galaxies,  baryon accretion in the center of the halo will suffer adiabatic contraction and drag the dark matter profile into an even steeper one,  and on the other hand, the baryonic characteristic of the gas causes it to radiate energy, increasing the entropy of the dark matter background and  preventing the contraction.  As our knowledge about these two processes is still incomplete, it is important  to take into account these effects in future works.

\begin{acknowledgements}
R.S.S.  thanks the Brazilian agencies FAPESP (2009/05176-4) and CNPq (200297/2010-4) for financial support. E.E.O.I. thanks the Brazilian agency CAPES (1313-10-0) for financial support.  We also thank the anonymous referee for fruitful comments and suggestions. This work was supported by World Premier International Research Center Initiative (WPI Initiative), MEXT, Japan.

\end{acknowledgements}

\end{document}